\documentstyle[amssymb,aps]{revtex}
\begin{document}
\title{Level correlations in integrable systems}
\author{R. A. Serota and J. M. A. S. P. Wickramasinghe}
\address{Department of Physics\\
University of Cincinnati\\
Cincinnati, OH\ 45221-0011\\
serota@physics.uc.edu}
\maketitle

\begin{abstract}
We derive a simple analytical expression for the level correlation function
of an integrable system. It accounts for both the lack of correlations at
smaller energy scales and for global rigidity (level number conservation) at
larger scales. We apply our results to a rectangle with incommensurate sides
and show that they are in excellent agreement with the limiting cases
established in the semiclassical theory of level rigidity.
\end{abstract}

\section{Introduction}

In the quantum limit, classically integrable systems are often described as
those with ''level bunching'' or, alternatively, absence of level
correlations, while classically chaotic ergodic systems are described as
exhibiting ''level rigidity''\cite{G}. We will argue here, however, that
such descriptions are overly simplistic and apply only in certain energy
intervals. In fact, since the total number of levels must be conserved, both
integrable and chaotic systems exhibit various degrees of rigidity depending
on the energy scales.

The main difference between the chaotic and integrable systems lies in the
relevant length and energy scales. Chaotic systems are characterized by
their diffusive behavior\cite{S1} and the relevant energy scales are the
mean level spacing $\Delta $, the Thouless energy $E_{c}=D/L^{2}$, and the
inverse scattering time $\tau ^{-1}=D/\ell ^{2}$, where $D\sim v_{F}\ell $
is the diffusion coefficient, $v_{F}$ is the Fermi velocity, $\ell $ is the
mean-free-path, and $L$ is the system size ($\hbar =c=1$ below). The
Thouless scale is associated with the diffusion to the system boundary and
can be written as\footnote{%
In what follows, we consider only 2D systems.} $E_{c}\sim \sqrt{\epsilon
_{F}\Delta }\left( \ell /L\right) $, where $\epsilon _{F}$ is the Fermi
energy.

For chaotic systems, on energy scales $E\ll $ $\Delta $, the level rigidity
statistics $\Delta _{3}$, which is based on the cumulative level
distribution function (see e.g.\cite{G}), is 
\begin{equation}
\Delta _{3}\sim \frac{E}{\Delta }  \label{rig_chao-1}
\end{equation}
indicating the lack of level correlations. The onset of rigidity occurs at $%
E\sim \Delta $ (and, hence, also known as level repulsion) and for $\Delta
\ll E\ll E_{c}$ 
\begin{equation}
\Delta _{3}\sim \ln \frac{E}{\Delta }  \label{rig_chao-2}
\end{equation}
This energy range corresponds to length scales between $L$ and $\sqrt{
D/\Delta }$ and the rigidity is related to the diffusive character of the
multiple traversals of the system. On scales such that $E_{c}\ll E\ll \tau
^{-1}$, which correspond to lengths between $\ell $ and $L$ and to
uncorrelated diffusion in different parts of the system, the level structure
is much less rigid\cite{AS}, 
\begin{equation}
\Delta _{3}\sim \frac{E}{E_{c}}\sim \frac{1}{g}\frac{E}{\Delta }
\label{rig_chao-3}
\end{equation}
where $g\sim E_{c}/\Delta \gg 1$ is the dimensionless conductance. The
existence of the above regimes is, nonetheless, in complete agreement with
the global rigidity (level number conservation) which can be expressed in
the following form: 
\begin{equation}
\int_{-\infty }^{\infty }d\omega \left\langle \delta \nu \left( \epsilon
\right) \delta \nu \left( \epsilon +\omega \right) \right\rangle =0
\label{rig_glob}
\end{equation}
where $\nu \left( \epsilon \right) $ is the level density, $\delta \nu
\left( \epsilon \right) =\nu \left( \epsilon \right) -\left\langle \nu
\right\rangle $ and $\left\langle \nu \right\rangle $ the mean level
density.(Here $\epsilon $ is the ''running'' value of energy which below
will be used interchangeably with $\epsilon _{F}$). This relationship has
been discussed, for instance in\cite{SS}, but without taking account of
diffusion modes. The diffusion modes are only known to be treated
perturbatively, but even then the above relationship is satisfied as can be
easily seen with the help of eq. (32) in \cite{AS}.

In integrable systems (\ref{rig_glob})\ must also be satisfied; otherwise
even the notion of the mean level density $\Delta $ would be dubious.
However, the only relevant scale in this case is 
\begin{equation}
\varepsilon \sim \sqrt{\epsilon _{F}\Delta }  \label{epsilon}
\end{equation}
In what follows, we shall derive the expression for the level correlation
function $\left\langle \delta \nu \left( \epsilon \right) \delta \nu \left(
\epsilon +\omega \right) \right\rangle $ in integrable systems that
satisfies (\ref{rig_glob}) and illustrate it on an example of a rectangle
with incommensurate sides (a generic integrable system). Applying this
expression to evaluation of the $\Delta _{3}$-statistics, we will find that
for $E\ll \varepsilon $ 
\begin{equation}
\Delta _{3}\sim \frac{E}{\Delta }\left( 1-O\left( \left( \frac{E}{%
\varepsilon }\right) ^{3}\right) \right)  \label{rig_inte-1}
\end{equation}
and for $E\gg \varepsilon $ 
\begin{equation}
\Delta _{3}\sim \frac{\varepsilon }{\Delta }\left( 1-O\left( \frac{%
\varepsilon ^{2}}{E^{2}}\right) \right)  \label{rig_inte-2}
\end{equation}
that is $E$-independent. The latter indicates that the spectrum becomes more
rigid at large scales, the fact obviously related to (\ref{rig_glob}). The
leading terms in (\ref{rig_inte-1}) and (\ref{rig_inte-2}) were a subject of
extensive numerical and analytical study in \cite{CCG} and \cite{B}.
However, the evaluation in the present paper of the level correlation
function over the entire energy range allows for the evaluation over the
entire energy range of $\Delta _{3}$- and $\Sigma $-statistics (and other
characterizations of the spectrum\cite{G}) also.

\section{Level correlation function}

We approach the derivation of the level-correlation function in two
different ways, which will later be shown to be equivalent. First, we
utilize the semiclassical formalism wherein the level density is expressed
as a sum over periodic orbits\cite{B},\cite{G}. In such a formalism, the
Fourier transform of the level correlation function is given by 
\begin{equation}
\phi \left( t\right) =\sum_{j}A_{j}^{2}\delta \left( t-T_{j}\right)
\label{phi1}
\end{equation}
where $A_{j}$ and $T_{j}$ are the periodic orbit amplitudes and periods
respectively\cite{B}. To simplify the calculation, we will express all
energies in terms of $\Delta $ so that $\Delta $ will be dropped below. It
is known that (see (58) in\cite{B}) 
\begin{equation}
\phi \left( t\rightarrow \infty \right) =1/2\pi  \label{phi-limit}
\end{equation}
and it is clear that $\phi \left( t\right) =0$ for $t<T_{\min }$, where $%
T_{\min }$ ($\sim \varepsilon ^{-1}$ in 2D) is the shortest periodic orbit.
Based on these limiting behaviors, we propose a very simple ansatz, namely, 
\begin{equation}
\phi \left( t\right) = 
\begin{array}{lll}
1/2\pi , &  & t>T_{\min } \\ 
0, &  & t<T_{\min }
\end{array}
\label{ansatz}
\end{equation}
Clearly, this ansatz is applicable in any dimension.

Since the correlation function should be symmetrical with respect to the $%
\omega \rightarrow -\omega $ transformation, the FT should be symmetrical
with respect to the $t\rightarrow -t$ transformation. The obvious
generalization of (\ref{phi1}) would be via the substitution $\delta \left(
t-T_{j}\right) \rightarrow \left[ \delta \left( t-T_{j}\right) +\delta
\left( t+T_{j}\right) \right] $ and the ansatz (\ref{ansatz}) is generalized
be means of

\begin{equation}
\phi \left( t\right) =\frac{1}{2\pi }\left[ 1+\frac{%
%TCIMACRO{\limfunc{sign} }
%BeginExpansion
\mathop{\rm sign}%
%EndExpansion
\left( t-T_{\min }\right) }{2}-\frac{%
%TCIMACRO{\limfunc{sign} }
%BeginExpansion
\mathop{\rm sign}%
%EndExpansion
\left( t+T_{\min }\right) }{2}\right]  \label{phi2}
\end{equation}
which satisfies $\phi \left( -t\right) =\phi \left( t\right) $. With such
definition, the correlation function becomes 
\begin{equation}
\left\langle \delta \nu \left( \epsilon \right) \delta \nu \left( \epsilon
+\omega \right) \right\rangle =\delta \left( \omega \right) -\frac{\sin
\left( \omega T_{\min }\right) }{\pi \omega }  \label{corr_fn}
\end{equation}

Consider now a rectangle whose sides $L_{1}$ and $L_{2}$ are such that 
\begin{equation}
L_{1}^{2}/L_{2}^{2}=\alpha   \label{alpha}
\end{equation}
is irrational. It is also assumed that $\alpha \lesssim 1$ (this assumption
is opposite to the assumption $\alpha \gtrsim 1$ in \cite{CCG},\cite{B}). We
also have 
\begin{equation}
\epsilon _{F}=N\Delta =\frac{2\pi }{mA}  \label{E_F}
\end{equation}
where $A=L_{1}L_{2}$ is the area and $N$ is the mean number of levels below
the Fermi level. With these definitions, the shortest periodic orbit is the
one with length $2L_{1}$ and 
\begin{equation}
T_{\min }=\frac{2L_{1}}{v_{F}}=\frac{2\pi ^{1/2}\alpha ^{1/4}}{\sqrt{%
\epsilon _{F}}}  \label{T_min}
\end{equation}
The level correlation function for this integrable system is then 
\begin{equation}
\left\langle \delta \nu \left( \epsilon \right) \delta \nu \left( \epsilon
+\omega \right) \right\rangle =\delta \left( \omega \right) -\frac{\sin
\left( 2\pi \omega /\varepsilon \right) }{\pi \omega }  \label{corr_fn-rect}
\end{equation}
with the definition 
\begin{equation}
\varepsilon =\sqrt{\pi \epsilon _{F}}/\alpha ^{1/4}  \label{epsilon2}
\end{equation}
(compare this definition with eqs. (7)\ and (43) in \cite{B}).

The alternative way of derivation is specific to a rectangle (since it is
based on the energy spectrum specific to the rectangular box) and involves a
slight modification of a derivation for a square in Appendix B of\cite{vO}.\
It is shown there that the level density can be written as 
\begin{equation}
\delta \nu \left( \epsilon \right) =\sqrt{\frac{2}{\pi }}\sum_{m_{1},m_{2}=-%
\infty }^{\infty }\frac{\cos \left( kL_{\sigma }\right) }{\left( kL_{\sigma
}\right) ^{1/2}}  \label{delta_nu}
\end{equation}
The term $m_{1}=m_{2}=0$ is excluded. (In fact, this term gives the mean
level density \cite{vO}). In (\ref{delta_nu}), $k=\sqrt{2m\epsilon }$, $%
kL\gg 1$ and 
\begin{equation}
L_{\sigma }=2\sqrt{\left( m_{1}L_{1}\right) ^{2}+\left( m_{2}L_{2}\right)
^{2}}  \label{L_sigma}
\end{equation}
Notice that reducing the summation to only positive $m_{1}$,$m_{2}$
demonstrates that (\ref{delta_nu}) is a summation over periodic orbits whose
length is given by (\ref{L_sigma}). Omitting the rapidly oscillating terms,
we find 
\begin{equation}
\left\langle \delta \nu \left( \epsilon \right) \delta \nu \left( \epsilon
+\omega \right) \right\rangle =\frac{1}{\pi }\sum \frac{\cos \left( \Delta
kL_{\sigma }\right) }{\left( kL_{\sigma }\right) ^{1/2}}
\label{corr_fn-rect2}
\end{equation}

The sum can be converted to an integral in polar coordinates, 
\begin{equation}
\left\langle \delta \nu \left( \epsilon \right) \delta \nu \left( \epsilon
+\omega \right) \right\rangle =\frac{1}{2kL_{1}L_{2}}\left[ \int_{-\infty
}^{\infty }\exp \left( 2i\Delta k\rho \right) d\rho -\int_{-\rho _{\min
}}^{\rho _{\min }}\cos (2\Delta k\rho )d\rho \right]  \label{corr_fn-rect3}
\end{equation}
where\footnote{%
More consistently, the variable of integration should be a dimensionless $%
L_{1}\kappa \ll 1$; introducing the latter, however, does not affect the
final result.} $\rho _{\min }\approx L_{1}$. The appearance of $\rho _{\min
} $ above is related to the fact that the term $m_{1}=m_{2}=0$ is excluded
in (\ref{delta_nu}). It has the meaning of the length of the shortest
periodic orbit, which is in direct relation to ansatz (\ref{phi2}). (Notice
that when $\alpha \sim 1$, $\rho _{\min }\approx L_{1}\approx L_{2}$ and the
combination of the cut-off and polar coordinates becomes a more tenable
approximation). Integrating (\ref{corr_fn-rect3}) and substituting $\Delta
k=\omega m/k$, we recover the previously obtained result (\ref{corr_fn-rect}%
).

\section{$\Delta _{3}$- and $\Sigma $-statistics}

We now proceed to apply (\ref{corr_fn-rect}) to the evaluation of $\Delta
_{3}$- and $\Sigma $-statistics. For $\Delta _{3}$-statistics, we begin with 
\begin{eqnarray}
\Delta _{3}\left( E\right) &=&2\int_{0}^{\infty }\frac{dt}{t^{2}}\phi \left(
t\right) G\left( Et/2\right)  \label{delta} \\
G\left( y\right) &=&1-F^{2}\left( y\right) -3\left( F^{\prime }\left(
y\right) \right) ^{2}\text{, }F\left( y\right) =\sin y/y  \nonumber
\end{eqnarray}
(compare with eq. (20) in\cite{B}). Substituting (\ref{phi2}) into (\ref
{delta}), we find 
\begin{equation}
\Delta _{3}\left( E\right) =\frac{E}{2\pi }\int_{ET_{\min }/2}^{\infty }dt%
\frac{G\left( t\right) }{t^{2}}  \label{delta2}
\end{equation}
This is easily evaluated analytically in terms of algebraic, trigonometric
and sine integral functions. However, of greater interest are the limiting
cases: 
\begin{equation}
\Delta _{3}\left( E\ll T_{\min }^{-1}\sim \varepsilon \right) \cong \frac{1}{%
15}E\left[ 1-\frac{\left( T_{\min }E\right) ^{3}}{144\pi }\right]
\label{delta_limit1}
\end{equation}
and 
\begin{equation}
\Delta _{3}\left( E\gg T_{\min }^{-1}\sim \varepsilon \right) \cong \frac{1}{%
\pi T_{\min }}\left[ 1-\frac{8}{3\left( T_{\min }E\right) ^{2}}\right]
\label{delta_limit2}
\end{equation}
For the rectangular, upon substitution of (\ref{T_min}), the first term in (%
\ref{delta_limit2}) becomes 
\begin{equation}
\Delta _{3}^{(rect)}\left( E\gg \varepsilon \right) =\frac{1}{2\pi
^{3/2}\alpha ^{1/4}}\sqrt{\epsilon _{F}}  \label{delta_limit2-rect}
\end{equation}
Remarkably, for $\alpha \sim 1$, the constant in front of the radical is
close ($\sim 5$ percent) to the value in (45) of\cite{B} which was obtained
by exact summation over all periodic orbits. This supports the utility of
the ansatz (\ref{phi2}) for the level correlation function.

For completeness we give the results for $\Sigma $-statistics as well. In
the same limiting cases, we find 
\begin{equation}
\Sigma \left( E\ll T_{\min }^{-1}\sim \varepsilon \right) \cong E\left[ 1-%
\frac{\left( T_{\min }E\right) }{\pi }\right]  \label{sigma_limit1}
\end{equation}
and 
\begin{equation}
\Sigma \left( E\gg T_{\min }^{-1}\sim \varepsilon \right) \cong \frac{2}{\pi
T_{\min }}\left[ 1+\frac{\sin \left( T_{\min }E\right) }{T_{\min }E}-\frac{%
2\cos \left( T_{\min }E\right) }{\left( T_{\min }E\right) ^{2}}\right]
\label{sigma_limit2}
\end{equation}
We observe that the accuracy of $\Delta _{3}$ in the limiting cases is
higher than that of $\Sigma $. This is because $\Delta _{3}$ describes the
cumulative behavior of the levels in the spectrum and is thus a more
appropriate characteristic for the rigidity. We also observe that the last
two terms in (\ref{sigma_limit2}) are rapidly oscillating. The most
interesting feature of both statistics for large energy intervals is that
the leading terms are $E$-independent (but do depend on $\epsilon $).

\section{Discussion}

The central result of this work is (\ref{corr_fn}). This simple form of the
correlation function is in excellent agreement with the limiting behavior of
the $\Delta _{3}$-statistics for both small and large energy scales and is
consistent with the overall spectral rigidity (\ref{rig_glob}). Having a
closed-form correlation function, applicable at all energy scales, allows
for the evaluation of the closed-form expressions for $\Delta _{3}$- and $%
\Sigma $-statistics as well, with the corrections in the appropriate regimes
given by eqs. (\ref{delta_limit1})-(\ref{delta_limit2}) and eqs. (\ref
{sigma_limit1})-(\ref{sigma_limit2}) respectively.

The significance of the corrections to the leading terms in (\ref
{delta_limit1}) and (\ref{sigma_limit1})\ is that they describe the onset of
level correlations at small scales, which are dominated by the $\delta $
-function in (\ref{corr_fn}) (corresponding to the absence of correlations
in the zeroth order). The corrections to the leading terms at large scales, (%
\ref{delta_limit2}) and (\ref{sigma_limit2}), are also of interest
especially in view of the issues related to the symmetry-breaking
perturbations (see below). For verification purposes, we are presently
working on the numerical evaluation of $\Sigma $-statistics which will be
reported elsewhere.

We emphasize that (\ref{corr_fn}) was obtained both via ansatz (\ref{phi2})
and, for the specific case of a rectangle with incommensurate sides, by a
direct evaluation in the energy space. In both circumstances the small
inaccuracy is related to the continuous nature of the approximation used
versus the discrete behavior in the limit of short orbits.

Clearly, if a term $\delta \phi \left( t\right) $, such that $\delta \phi
\left( T<T_{\min }^{\prime }\right) =0$ and $\delta \phi \left( \pm \infty %
\right) =0$, is added to (\ref{phi2}), eq. (\ref{rig_glob}) will still be
satisfied and the leading terms in (\ref{delta_limit1}) and (\ref
{sigma_limit1}), determined by the $\delta $-function in (\ref{corr_fn}),
will remain unchanged. Furthermore, since the relevant scales are $T_{\min
}^{\prime }\gtrsim T_{\min }$, the corrections introduced at large energy
scales will have the same functional form as, and will be of the order of, (%
\ref{delta_limit2}) and (\ref{sigma_limit2}). The question of $\delta \phi
\left( t\right) $ may be relevant in discussion of the effect of
time-reversal symmetry-breaking terms, such as magnetic field. We hope to
address this question in a future publication.

\section{Acknowledgments}

We are grateful to B. Goodman for many helpful discussions and for careful
reading of the manuscript.

\end{document}